\documentclass[acmtog]{acmart}

\usepackage{booktabs} 

\usepackage{wrapfig}
\usepackage{mathtools}
\usepackage{enumitem}
\usepackage{siunitx}
\usepackage[ruled]{algorithm2e} 

\SetAlFnt{\small}
\SetAlCapFnt{\small}
\SetAlCapNameFnt{\small}
\SetAlCapHSkip{0pt}

\makeatletter
\@ACM@balancefalse
\makeatother

\acmJournal{TOG}

\begin{document}
\title{Convex Collision-Free Regions}

\author{Tomoyo Kikuchi}
\affiliation{%
 \institution{Kyoto University}
 \country{Japan}
}
\author{Takashi Kanai}
\affiliation{%
 \institution{The University of Tokyo}
 \country{Japan}
}


\begin{abstract}
\emph{Convex Collision-Free Regions} (CCFR) is a collision handling method that explicitly represents local convex feasible regions to enforce non-penetration.
Each feasible region is constructed from surrounding mesh primitive configurations, including edge-edge and vertex-face interactions. The resulting convex region represents admissible non-penetrating vertex displacements at the current configuration.
Existing collision handling methods for deformable body simulation have largely relied on implicit representations of feasibility, resulting in either compromised robustness for secondary collisions and codimensional contacts or tight coupling with specific nonlinear optimization schemes.
Our formulation constructs feasible regions independently for each vertex, defined prior to penetration, inherently accounts not only for primary collisions but also for secondary collisions and codimensional contacts, enabling highly scalable and parallelizable collision handling.
These feasible regions encode geometric non-penetration constraints independently of physical contact response models.
CCFR does not rely on nonlinear optimization and is compatible with simulation frameworks such as Extended Position-Based Dynamics (XPBD) that do not explicitly maintain interior feasibility during iterative updates.
The effectiveness of CCFR is demonstrated across cloth, hair, wire, particle systems, and codimensional contact scenarios, showing versatile and efficient collision handling.
\end{abstract}

%
%

%
%


\begin{teaserfigure}
  \includegraphics[width=\textwidth]{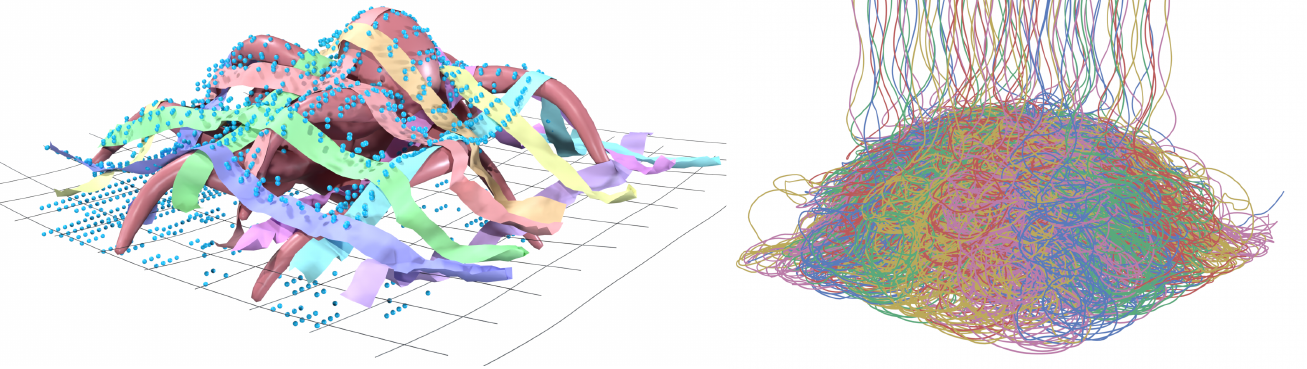}
  \caption{Collision handling results produced by the proposed Convex Collision-Free Regions (CCFR).}
  \Description{Collision handling results produced by the proposed CCFR.}
  \label{fig:teaser}
\end{teaserfigure}

\maketitle

\section{Introduction}
Collision handling is one of the most fundamental yet challenging problems in physics-based animation and simulation.
Accurately preventing interpenetration between geometric primitives is essential not only for physical plausibility, but also for the robustness and efficiency of numerical solvers in practical simulation systems.

The non-penetration condition requires that geometric primitives do not interpenetrate, and is commonly expressed as an inequality constraint stating that the distance between two interacting primitives must be non-negative.
Formally, a non-penetration constraint can be written as:
\[
\phi_i(\mathbf{x}) \ge 0,
\]
where $\phi_i(\mathbf{x})$ denotes a constraint function measuring separation between interacting primitives associated with the $i$-th potential contact.
Under this formulation, collision handling can be viewed as an optimization problem subject to inequality constraints:
\[
(P) \quad
\begin{cases}
\text{minimize} & f(\mathbf{x}), \\
\text{subject to} & \phi_i(\mathbf{x}) \ge 0, \quad \forall i .
\end{cases}
\]
A configuration $\mathbf{x}$ that satisfies all non-penetration constraints is called a feasible solution, and the set of all such configurations is referred to as the feasible region.
Contact and friction are defined on the boundary of the feasible region.
While rigid-body simulations often resolve these via explicitly defined feasible regions and complementarity problems, extending such global feasible-region constructions to deformable bodies with complex geometry or a large number of interacting primitives becomes prohibitively difficult.

Consequently, collision handling for deformable bodies has largely relied on implicit representations of feasibility.
As a result, existing formulations do not distinguish between geometric
corrections required to enforce non-penetration and physical contact responses
such as repulsion and friction.
This conflation is particularly problematic for thin structures, leading to secondary collisions after an initial correction.
To address this issue, fail-safe methods \cite{Provot97, Bridson02, Huh01, Harmon08} and post-processing untangling methods~\cite{Baraff03, Volino06, Wicke06, Ye12} have been proposed; however, they do not always guarantee complete resolution of collisions.
Methods resolving collisions only after penetration occurs \cite{Macklin20SDF, Chen23, Allard10, Kikuchi25} are widely adopted in practice due to their conceptual simplicity; however, interpenetration must occur before corrective forces are generated. 
As a result, they struggle to prevent intersections in advance, limiting their generality in complex scenarios such as cloth self-collisions.
Barrier-based methods \cite{Li20, Li21} have been proposed to improve the robustness of collision handling across a wider range of contact scenarios, and have been adopted in computer graphics.
In these methods, the barrier energy for the penetrated state $\phi_i \leq 0$ is deliberately left undefined, since allowing such states would explicitly admit interpenetration.
Instead, the formulation enforces that all iterates remain strictly inside the penetration-free region, which inherently ties the collision handling mechanism to an interior-point-style optimization process and limits the range of admissible simulation schemes.
In addition, numerical robustness is typically achieved by activating repulsive forces within a finite distance from contact.
As this activation range is enlarged, primitives that would not collide in the exact penetration-free configuration are nevertheless subjected to artificial repulsion, making the geometric boundary of the feasible region ambiguous or ill-defined.

In this work, we introduce \emph{Convex Collision-Free Regions} (CCFR), a method that explicitly defines local feasible regions for large-scale and complex collision scenarios, as shown in Figure \ref{fig:teaser}.
The regions are constructed from surrounding mesh primitive interactions (VF, EE, FV), and represented locally per vertex for scalability and parallelization.
This representation naturally accommodates not only primary collisions but also secondary collisions and codimensional contacts.
The explicit representation of the feasible region decouples geometric non-penetration constraints from physical contact responses. 
As a result, non-penetration is enforced solely through projection onto the convex feasible
region, while repulsion and friction are activated only as physical responses
within the activation band.
The proposed method eliminates the need for Hessian evaluation and stiffness tuning typically required in optimization-based formulations. This makes the proposed method particularly well suited for integration into fast position-based frameworks such as Extended Position-Based Dynamics (XPBD) \cite{Macklin16}, while preserving their efficiency and implementation simplicity.

In summary, our contributions are:
\begin{itemize}
    \item A collision formulation that explicitly represents local feasible regions for non-penetration, inherently accounting for secondary collisions and codimensional contacts.
  \item An explicit separation between geometric non-penetration constraints and physical contact response models through locally defined feasible regions. 
  \item A per-vertex formulation that enables efficient parallelization and scalability.
  \item A projection scheme that enforces non-penetration without nonlinear optimization.
\end{itemize}

\section{Related Work}\label{sec:relwork}

\subsection{Continuous Approaches to Penetration-Free Motion}

Continuous collision detection (CCD) prevents interpenetration by accounting for the continuous motion of objects, thereby avoiding tunneling artifacts inherent to discrete collision detection (DCD).
Early CCD methods \cite{Moore88, Provot97,Brochu12,Tang14,Wang15,Wang22} aimed to compute the exact time of impact (TOI), defined as the earliest time at which the distance between two primitives becomes zero.
Examples of primitives include vertex--vertex, vertex--edge, vertex--face, edge--edge, edge--face, and face--face.

Historically, the TOI was primarily used to regulate the time step size $\Delta t$, indicating how far the simulation could safely advance in time before a collision occurs.
In this sense, early CCD methods implicitly defined a feasible region in the time domain, in contrast to the spatial feasible regions discussed in the previous sections.
However, although theoretically precise, such formulations require solving nonlinear equations and performing repeated geometric overlap tests, making them numerically fragile and computationally expensive for complex deformable models.
In cloth simulation, it is desirable to relax the definition of TOI from the moment when the distance becomes zero to a formulation that incorporates thickness (i.e., surface offsets).
Accordingly, conservative CCD methods that enforce a small but guaranteed separation distance have been proposed to more reliably prevent interpenetration~\cite{Harmon11, Lu19}.
A representative example is Additive CCD (ACCD)~\cite{Li21}, which iteratively advances colliding primitives and terminates the process before full intersection occurs.
While this strategy improves robustness, it still relies on repeated distance queries and updates within a single time step.

In position-based optimization frameworks, the state is updated directly in configuration space, and the role of TOI shifts accordingly.
Rather than controlling the time step, TOI is reinterpreted as a bound on the admissible spatial displacement, often serving as a step size $\Delta d$ constraint in line search procedures \cite{Li20}.
As a result, the notion of feasibility gradually transitions from the temporal domain to spatial feasible regions defined in the space of positions.
This perspective naturally leads to distance-threshold-based formulations, which 
enforce penetration-free states by directly constraining the admissible spatial feasible region of motion.
For example, Shen et al. \shortcite{Shen24} bound the step size using a global repulsion activation threshold.
Chen et al. \shortcite{Chen25} employ the penetration-free simulation technique proposed by Wu et al. \shortcite{Wu20}, in which each vertex is constrained by a conservative bound determined from the distance to its closest point.
Both approaches maintain penetration-free states by directly enforcing distance-threshold-based constraints.
As a result, non-penetration is inherently guaranteed by these constraints and can be verified at the current configuration using DCD, eliminating the need for CCD or iterative TOI estimation.

\subsection{Maintaining the Feasible Region}
\label{sec:rel-work-3}
The feasible region is defined as the intersection of multiple linear half-spaces, whose boundaries form a piecewise-linear constraint manifold.
Historically, such convex regions have been handled using explicit geometric representations,
for example via polytope construction \cite{barber96quickhull} or geometric feature-tracking
methods such as V-Clip \cite{Mirtich98} and the Lin-Canny algorithm \cite{Lin91}.
While these approaches are highly efficient for static or coherently moving shapes,
they rely on maintaining explicit boundary or Voronoi structures.
In dynamic mesh simulations, where the underlying half-space definitions change continuously,
the cost of updating these geometric structures can become prohibitive.

Given the limitations of explicit geometric methods, algebraic approaches provide a more robust framework. 
For general background on optimization-based approaches to constrained problems,
we refer the reader to the textbook by Nocedal and Wright \cite{nocedal06numerical}.
When enforcing constraints through projection or optimization, the local behavior of the feasible region is governed by the active boundaries; consequently, computing an optimal or projected position reduces locally to solving a system of linear equalities.
Classical direct solvers for linear systems, such as Gaussian elimination, QR decomposition, or singular value decomposition (SVD) \cite{golub13, meyer23}, are highly effective when the feasible region is defined by a fixed,
minimal, and internally consistent set of constraints that directly yields a well-posed system of equations. 
However, in dynamic simulations, the feasible region is determined by a collection of candidate constraints generated on the fly.
This collection often includes redundant, inactive, or even mutually incompatible constraints,
and the specific subset that defines the boundary of the feasible region is not known \textit{a priori}.
As a result, the constraint system cannot be readily assembled into a stable linear system suitable for direct numerical methods.

\begin{wrapfigure}[7]{r}{0.3\columnwidth}
  \centering
  \vspace{-0.8\intextsep}
  \includegraphics[width=\linewidth]{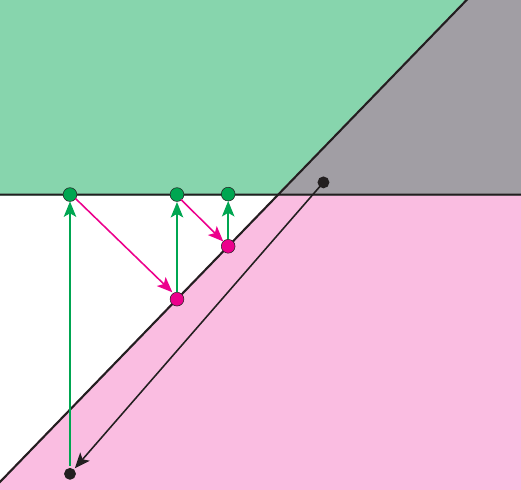}
\end{wrapfigure}

An alternative algebraic approach involves projection-based methods, which aim to map a point violating constraints back onto the feasible region by finding the closest point.
Among these, the successive projection method \cite{BREGMAN67} and the active-set-based methods are particularly attractive in dynamic settings, as they do not require the active constraint set to be known in advance and can operate incrementally as constraints are generated.
However, such iterative corrections do not guarantee that previously satisfied constraints remain satisfied, which can suffer from slow convergence or ``zigzagging'' behavior (see inset), particularly when the convex region forms sharp angles \cite{gill19}. 

To address these limitations, Plesn\'{i}k \shortcite{Plesnik07} proposed an algebraic projection method
that is particularly well suited to settings where the active constraint set is not known in advance.
Rather than solving a fixed system of equations, the method incrementally constructs an affine subspace
spanned by the currently active hyperplanes, while ensuring that all previously satisfied constraints
remain satisfied.
This property leads to finite-step convergence and effectively suppresses the oscillatory behavior
observed in successive projection methods, especially in sharp-angled feasible regions.

\section{Method}
\label{sec:method}

This section presents a collision handling formulation based on an explicit geometric characterization of non-penetration and physical contact responses.
The core of the proposed approach is to decompose collision handling into three distinct stages:
(i) defining a local convex feasible region encoding non-penetration constraints,
(ii) enforcing feasibility through projection-based updates, and
(iii) integrating repulsion and friction responses within XPBD independently of the geometric feasibility constraints.


This approach is referred to as \emph{Convex Collision-Free Regions} (CCFR), as non-penetration and contact admissibility are represented by a convex feasible region whose boundary encodes all geometrically valid contact configurations.
Since the feasible region is defined locally for each particle based on its
non-penetration constraints, collision handling can be performed entirely via local projection steps, thereby avoiding the construction or solution of a global constraint system.

\subsection{Definition of Convex Feasible Region}\label{subsec:feasible-region}
The condition of non-penetration is fundamentally defined by the requirement that the distance between any two particles must remain non-negative: 
\[ 
    \phi(\mathbf{x}_\mathrm{A}, \mathbf{x}_\mathrm{B}) \geq 0.
\] 
However, when collision handling is performed via iterative local projections, the direct enforcement of such pairwise geometric constraints can lead to order-dependent updates and introduce secondary collisions.
To address this limitation, the non-penetration condition is reformulated as a set of linear constraints on the allowable updated position of each particle, defined independently.
 
Let $\mathbf{x}_\mathrm{A}$ and $\mathbf{x}_\mathrm{B}$ denote the current positions of two particles, and let the unit separation direction be defined as:
\begin{equation}\label{eq:contact-normal}
    \mathbf{n} = \frac{\mathbf{x}_\mathrm{A} - \mathbf{x}_\mathrm{B}}
{\|\mathbf{x}_\mathrm{A} - \mathbf{x}_\mathrm{B}\|}.
\end{equation}
For the updated positions
$\mathbf{x}_\mathrm{A}' = \mathbf{x}_\mathrm{A} + \Delta \mathbf{x}_\mathrm{A}$
and
$\mathbf{x}_\mathrm{B}' = \mathbf{x}_\mathrm{B} + \Delta \mathbf{x}_\mathrm{B}$,
the non-penetration condition projected along $\mathbf{n}$ is expressed as:
\[ 
    \phi(\mathbf{x}_\mathrm{A}, \mathbf{x}_\mathrm{B}) \coloneqq 
\mathbf{n}^\top (\mathbf{x}_\mathrm{A}' - \mathbf{x}_\mathrm{B}') 
\ge 0.
\] 
Solving this inequality for $\mathbf{x}_\mathrm{A}'$ yields:
$ 
    \mathbf{n}^\top \mathbf{x}_\mathrm{A}' \geq \mathbf{n}^\top \mathbf{x}_\mathrm{B} + \mathbf{n}^\top \Delta \mathbf{x}_\mathrm{B}.
$ 
%
Since the displacement $\Delta \mathbf{x}_\mathrm{B}$ is generally unknown due to internal constraints and external forces---and may even be directed toward $\mathbf{x}_\mathrm{A}$ by contacts with other primitives---a conservative, mass-weighted bound independent of $\Delta \mathbf{x}_\mathrm{B}$ is imposed on the updated position. 
Incorporating an offset $\xi$, which accounts for the cloth thickness and defines
a minimum separation distance between contacting primitives, yields the following linear inequality constraint on $\mathbf{x}_\mathrm{A}'$:
\begin{equation}\label{eq:dpa3}
    \mathbf{n}^\top \mathbf{x}_\mathrm{A}' \geq \mathbf{n}^\top \mathbf{x}_\mathrm{A} + s\,\left(\mathbf{n}^\top (\mathbf{x}_\mathrm{B} - \mathbf{x}_\mathrm{A}) - \xi\right).
\end{equation}
where $s = \frac{w_{\mathrm{A}}}
{w_{\mathrm{A}} + w_{\mathrm{B}}}$ with $w_{\mathrm{A}}$ and $w_{\mathrm{B}}$ representing the inverse masses.

For a discretized object representation, analogous inequalities are established for each vertex with respect to the closest points on proximal primitives (vertices, edges, and faces).
Each inequality defines a half-space in the position space.
The feasible region $\mathcal{K}$ is therefore defined as the intersection of all such half-spaces, thus forming a convex polyhedron:
\begin{equation}\label{eq:convex-polyhedron}
    \mathcal{K} = \bigcap_{c \in \mathcal{C}} \left\{ \mathbf{x}' \mid g_c(\mathbf{x}') \geq 0 \right\}, \quad g_c(\mathbf{x}') = \mathbf{n}_c^{\top} \mathbf{x}' - b_c,
\end{equation}
where $\mathcal{C}$ denotes the set of all active constraints $c$. For each constraint $c$, $\mathbf{x}$ represents the current position of the constrained vertex, $\mathbf{x}_{\mathrm{B}_c}$ is the closest point on the opposing primitive, and the scalar term is given by $b_c = \mathbf{n}_c^\top \mathbf{x} + s_c\,\left(\mathbf{n}_c^\top (\mathbf{x}_{\mathrm{B}_c} - \mathbf{x}) - \xi\right)$.
It is worth noting that the feasible regions defined for different vertices at the same time step do not overlap: each region represents a disjoint, penetration-free subset of the displacement space associated with a single vertex.
This property is analogous to the safe domain implicitly enforced by the barrier functions in IPC \cite{Li20} and OGC \cite{Chen25}, and constitutes one of the key factors underlying the robustness of the proposed formulation.

This convex feasible region enables the formulation of collision avoidance as a constrained optimization problem. 
Defining the global updated position vector as $x' \coloneqq (\mathbf{x}_0', \mathbf{x}_1', \dots, \mathbf{x}_{n-1}')$, and $x, \Delta x$ similarly, the problem is stated as follows:
\begin{equation}
    \begin{cases}
    \text{minimize}_{\Delta x} & f(x'), \quad x' = x + \Delta x, \\[3pt]
    \text{subject to} & x_i' \in \mathcal{K}_i, \quad x_i' \in \mathbb{R}^3,
    \end{cases}
    \label{eq:minimization}
\end{equation}
The objective function $f(x')$ typically represents elastic potential energy,
an external force potential, or a deviation from an unconstrained candidate
solution.

\begin{figure}
  \centering
  \includegraphics[width=.5\columnwidth]{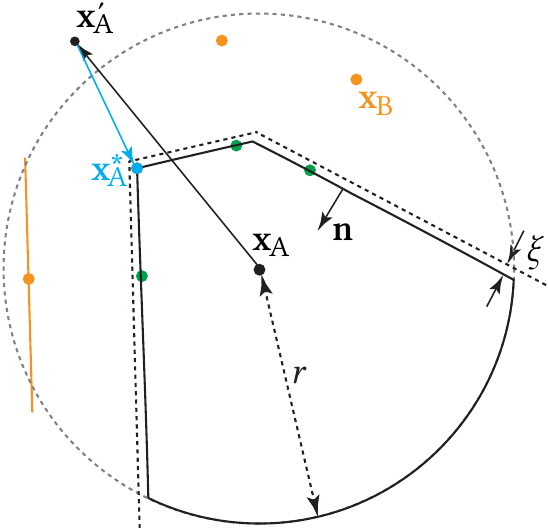}
  \caption{2D conceptual visualization of the local convex feasible region for a particle (actual constraints are defined in 3D).
  The region (bounded by solid black lines) is defined for a query vertex $\mathbf{x}_\mathrm{A}$ (black point) relative to the closest points $\mathbf{x}_\mathrm{B}$ (orange points) on proximal primitives (vertices, edges, or faces) within a search radius $r$ (dotted circle). 
  Each boundary line corresponds to a contact-induced half-space with inward normal $\mathbf{n}$ and offset $\xi$, representing the minimum separation distance.
  This configuration identifies the closest point (green point) on the constraint boundary.
  The black arrow originating from $\mathbf{x}_\mathrm{A}$ denotes the candidate displacement induced by internal constraints and external forces prior to projection.
  The resulting projected position $\mathbf{x}_\mathrm{A}^*$ (blue point) is obtained by projecting the candidate position $\mathbf{x}_\mathrm{A}'$ onto the convex feasible region using Plesn\'ik's algorithm.
  }
  \label{fig:convex}
\end{figure}

To ensure scalability, constraint evaluation is restricted to primitive pairs within a local search radius $r$, Each primitive contributes a local neighborhood of radius $r$; consequently, interactions are considered when the closest-point distance between primitive pairs falls within an effective range of $2r$. The feasible region therefore remains spatially local while avoiding unnecessary global collision queries.
The feasible region for a particle is illustrated in Figure~\ref{fig:convex}.
The same construction also applies when the closest interacting point lies on an edge or triangle primitive; in such cases, the resulting feasible constraints are distributed to the associated vertices and represented locally on a per-vertex basis.
Ideally, $r$ corresponds to the maximum expected displacement over the time step $\Delta t$;
however, as this involves a trade-off between physical accuracy and computational efficiency, $r$ is treated as a user-defined parameter in practice.
Furthermore, let $N_c$ denote the number of constraints identified within this local region. 
To circumvent the $\mathcal{O}(N_c^2)$ complexity associated with the explicit construction of the convex polyhedron $\mathcal{K}$---i.e., finding all intersections of $N_c$ half-spaces---neither $\mathcal{K}$ nor the intersection connectivity is explicitly stored.
Instead, constraints are dynamically evaluated using the method detailed in Section \ref{subsec:projection}, which yields a solution equivalent to an explicit projection onto the convex polyhedron $\mathcal{K}$ defined by these constraints.
%
%

\subsection{Projection onto the Feasible Region Boundary}\label{subsec:projection}
During the simulation, any excursion of a vertex outside this convex region indicates a violation of the non-penetration constraint.
Consequently, to strictly enforce collision avoidance, such an infeasible vertex must be projected onto the closest point $\mathbf{x}^*$ on the boundary of the convex region.
Since the convex region is defined as the intersection of half-spaces from Equation \eqref{eq:dpa3}, 
this boundary is formed by the set of violated constraints satisfied with equality:
\[
\mathbf{n}^\top \mathbf{x}^* = \mathbf{n}^\top \mathbf{x}_\mathrm{A} + s\,\left(\mathbf{n}^\top (\mathbf{x}_\mathrm{B} - \mathbf{x}_\mathrm{A})  - \xi\right).
\]
Locating $\mathbf{x}^*$ is equivalent to finding the orthogonal projection onto the solution set of the system of these linear equations, as discussed in Section \ref{sec:rel-work-3}. 
For this purpose, we adopt the method proposed by Plesn\'{i}k~\shortcite{Plesnik07}, which balances computational efficiency with robustness for large-scale, time-varying primitives. This projection process is illustrated by the blue arrow in Figure \ref{fig:convex}; for further algorithmic details, the reader is referred to the appendix.

The proposed projection independently modifies vertex displacements with respect to predicted positions; therefore, strict momentum conservation is generally not guaranteed.
In practice, nearby interactions are partially coordinated through repulsive responses activated prior to penetration, allowing neighboring vertices to influence each other before strong geometric projection occurs.
As the projection is applied independently of internal elastic forces, excessively large displacement truncations may introduce inconsistencies with the underlying elastic motion.
To mitigate this issue, smaller timesteps are preferred, as they limit per-step displacement relative to the feasible region boundary.
Moreover, the local per-vertex nature of the proposed method makes smaller timesteps particularly effective at avoiding situations in which only a subset of vertices become strongly constrained.


\paragraph{Numerical Treatment}
The original method in \cite{Plesnik07} is predicated on exact arithmetic; therefore, in a practical implementation, numerical errors must be accounted for.
To this end, a minor retraction mechanism is introduced: the projected boundary position $\mathbf{x}^*$ is shifted backward by an infinitesimal margin $\varepsilon$ (e.g., $10^{-6}$) along the displacement vector $(\mathbf{x}^* - \mathbf{x})$.
This $\varepsilon$ serves purely as a numerical tolerance to mitigate floating-point inaccuracies and does not represent a physical parameter; it is selected to be effectively negligible relative to the geometric scales of the simulation. A single fixed value of $\varepsilon$ is employed across all experiments, requiring no per-scene or per-example tuning.
Specifically, the projected boundary position $\mathbf{x}^*$ is adjusted as:
\[
\mathbf{x}^* \gets \mathbf{x} + \frac{\|\mathbf{x}^* - \mathbf{x}\| - \varepsilon}{\|\mathbf{x}^* - \mathbf{x}\|}\,(\mathbf{x}^* - \mathbf{x}) .
\]
However, if the magnitude of the displacement is less than $\varepsilon$, the movement is treated as negligible, and the position is reset to $\mathbf{x}^* = \mathbf{x}$.
Furthermore, if the algorithm fails to converge within the user-specified number of iterations, it is highly likely that the constraints are extremely tight or degenerate (i.e., nearly linearly dependent). In such cases, the position is conservatively reset to $\mathbf{x}^* = \mathbf{x}$ as a practical fallback, temporarily halting the vertex motion.
In practice, persistent halting is rarely observed because contact configurations continuously evolve through subsequent iterations and timesteps.
As neighboring constraints change, previously stalled vertices typically recover admissible displacement directions and resume motion in later updates.
Even under challenging dense-contact scenarios, we did not observe long-term locking caused by repeated halt operations as shown in  Figure \ref{fig:num-halting}.

\begin{figure}[tb]
  \centering
  \includegraphics[width=\columnwidth]{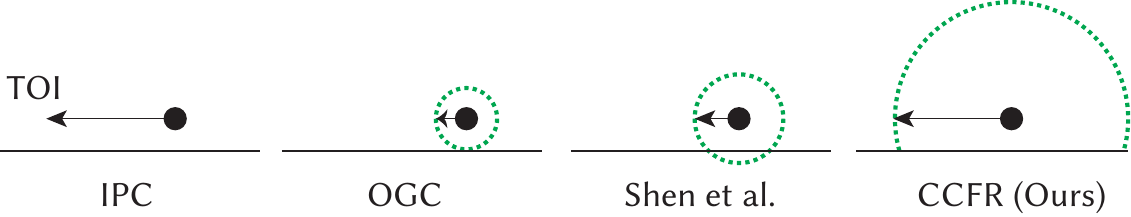}
  \caption{Comparison of penetration-free constraint formulations within a single solver iteration. Post-constraint velocities are denoted by arrows, and the admissible regions defined by each method are outlined by green dashed boundaries. IPC limits vertex advancement strictly to the TOI, whereas OGC restricts motion within a distance-based trust region. The method of Shen et al. advances all primitives by a conservative fraction of a global distance threshold. In contrast, CCFR provides a convex feasible region based on a user-defined displacement bound $r$, effectively preserving the intended direction of motion while strictly enforcing non-penetration.}
  \label{fig:comp-penetration-free}
\end{figure}

\paragraph{Comparison with CCD- or Distance-based Constraints}
Several representative schemes for preserving penetration-free motion within a single solver iteration are compared, focusing on how each method restricts the admissible displacement of a vertex.
Figure~\ref{fig:comp-penetration-free} illustrates a simple scenario in which a vertex moves along a flat floor, with the velocity direction indicated by the arrow.
This example isolates the displacement-limiting behavior of each method,
allowing a direct comparison of their admissible vertex motions without
being dominated by barrier-based line-search procedures.
IPC \cite{Li20} enforces non-penetration using a CCD-aware line search, limiting vertex advancement strictly to the time of impact (TOI).
As a result, the admissible displacement is bounded by the earliest collision event along the trajectory, often resulting in a conservative update.
In OGC \cite{Chen25}, the admissible motion is restricted by a distance-based trust region.
In this example, the proximity to the floor dictates the radius of the trust region, constraining the vertex to remain within the corresponding spherical bound.
Although this formulation is independent of the velocity direction, the resulting motion can be overly restricted when the closest constraint is orthogonal to the intended direction of motion.
The method of Shen et al.~\shortcite{Shen24} advances all primitives by at most
a conservative fraction of a global repulsion distance threshold.
While this guarantees robustness, the admissible displacement is uniformly limited and does not adapt to the local geometric context.

In contrast, the proposed CCFR method defines a user-specified displacement bound $r$ as an initial search region, from which a convex feasible region is carved out by constraints induced by the floor.
Crucially, this geometric restriction does not interfere with the intended direction of motion.
As a result, the vertex is permitted to advance along the floor while maintaining non-penetration, avoiding unnecessary constraints from contacts that are irrelevant to the direction of motion.


\subsection{Response and Friction Models in XPBD}\label{subsec:local-equality}



While various response models can be employed, a simple quadratic response model is adopted as a concrete realization for integrating the proposed formulation into XPBD.
To achieve smooth near-contact response behavior, a soft response penalty is introduced.
Let $d_c$ denote the signed distance evaluated along the contact normal,
and let $\delta$ be a user-defined repulsion range.
The response energy is defined as
\[
E_c^{\mathrm{resp}}
=
\frac{k}{2}
\,
w_c
\,
\max\!\left(0,\, \delta - d_c \right)^2,
\]
where $k$ is the stiffness parameter and
$w_c \in [0,1]$ is a visibility weight that attenuates
occluded constraints.
To prioritize the innermost contacts and suppress redundant farther responses,
each constraint is assigned a visibility weight.
For a reference constraint $c$ and a neighboring constraint $j$,
we define the occlusion factor
$
o_{cj}
=
S\!\left(
\frac{d_c - d_j}{\delta_v}
\right),
$
where $\delta_v$ controls the smooth transition width and
$S(\cdot)$ denotes a cubic smoothstep function.
Constraints with smaller distances therefore produce stronger occlusion of farther constraints.
The effective visibility weight is accumulated multiplicatively as
$
w_c
=
\prod_{j \ne c}
(1 - o_{cj}),
$
yielding reduced responses for constraints that are largely occluded by closer contacts.
This suppresses redundant response forces while preserving the smooth activation behavior of the soft penalty formulation.

Within the XPBD framework, the response energy can be equivalently reformulated as a position-level equality constraint.
For each active contact $c$, we define the response constraint
\[
C_c^{\text{resp}}(\mathbf{x}') = \max\!\left(0, \delta - d_c\right).
\]
The stiffness is controlled through the XPBD compliance parameter $\alpha = 1/k$, which governs
the accumulation of the corresponding Lagrange multiplier.

Friction is applied to the same set of physically active contacts identified for repulsion. 
Specifically, a tangential friction model is employed, following the approach of Macklin et al.~\shortcite{macklin14}.

\subsection{Parallelization}\label{subsec:graph-coloring}
The proposed formulation presents a significant advantage for parallel processing.
In standard PBD/XPBD frameworks, constraints are processed iteratively, and collision handling introduces constraints defined over collision primitives.
Since these collision primitives and their connectivity change dynamically at every collision detection step, precomputed graph coloring cannot be directly applied.

In contrast, the proposed CCFR method maintains constraints that are consistently defined on a per-vertex basis.
Although the constraints are updated during collision detection,
their ownership remains fixed at the vertex level.
Consequently, collision constraints can be processed in parallel over vertices without requiring graph coloring to be recomputed for each collision update.

\section{Results}\label{sec:results}

\subsection{Implementation Details}
During each simulation step, we first use AABB-trees in the broad phase to identify potentially intersecting triangle pairs. The AABBs are expanded by $(r + 0.5 \xi + 0.5 \delta)$. 
Next, we compute the shortest vertex--face and edge--edge distances of each triangle pair with the culling method proposed by Curtis~\shortcite{Curtis08} to determine the convex constraints.
Primitive pairs that are already intersecting in the initial configuration are excluded from this process.
Finally, the proposed CCFR method is applied to resolve the infeasible points.
These collision detection, constraint evaluation, and response steps are performed at every solver iteration.
All collision processing in our implementation is based on DCD,
since CCFR explicitly defines a local feasible region from the current configuration
and resolves collisions by projection.
The timing results are measured on an Intel Core i9-11900 CPU (8 cores, 16 threads) using CPU-based parallelization.
The parameter details for all experiments are provided in the appendix.
Our complete implementation of the CCFR method will be made available via a public GitHub repository.


\subsection{Examples}\label{subsec:fundamental-tests}

%

\begin{figure}
  \centering
  \includegraphics[width=.8\columnwidth]{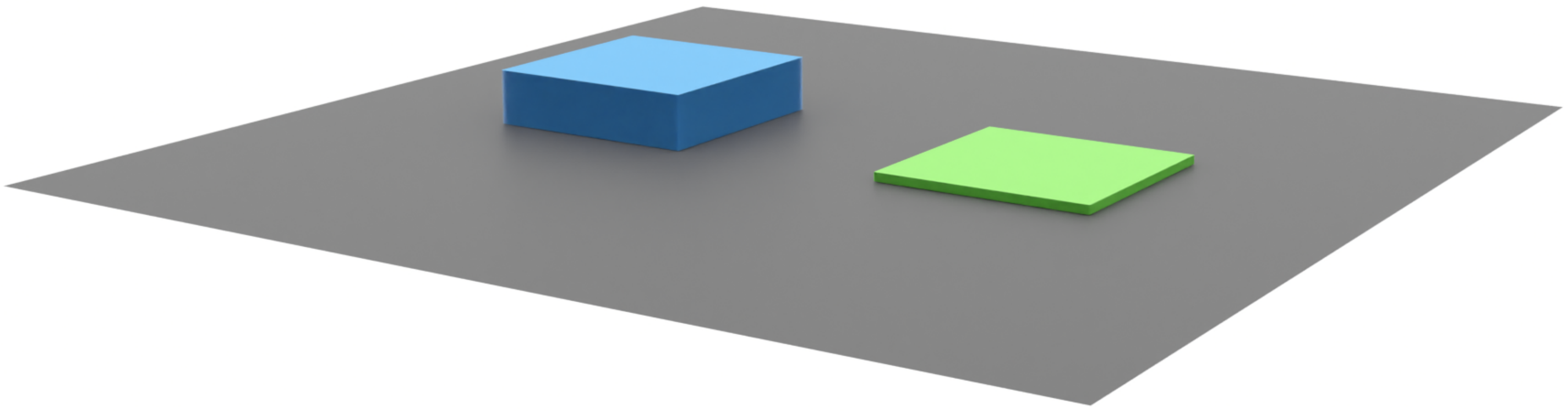}
  \caption{Examples of card stacks. The offset distance is set to 0.01 for the blue stack (left) and 0.001 for the green stack (right).}
  \label{fig:cards}
\end{figure}

\noindent\textbf{Degeneration Test.}
We compare the effect of the contact offset distance parameter $\xi$ on the visual thickness of card stacks, each consisting of 52 cards. The offset distance was set to $0.01$ for the blue stack and $0.001$ for the green stack (see Figure~\ref{fig:cards}). Although the simulation starts from an initial static state where the cards are already stacked, meaning the convex constraints are degenerate, all stacks maintain static equilibrium even under gravity. 

\begin{figure}
  \centering
  \includegraphics[width=\columnwidth]{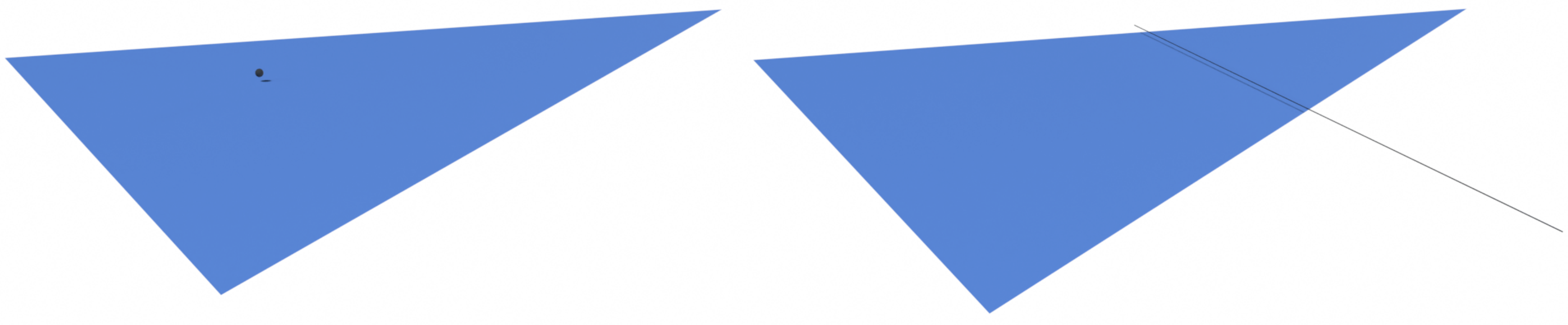}
  \caption{Examples of collision between the large triangle and a point (left), and collision between the large triangle and an edge (right) with an exaggerated offset for clarity. .}
  \label{fig:point-large}
\end{figure}

\noindent\textbf{Large Triangle Test.}
Figure~\ref{fig:point-large} shows collision handling between a point or an edge and a triangle larger than the radius $r$. Even when all distances to the triangle vertices exceed $r$, penetration is successfully prevented as long as the distance to the closest point on the triangle remains small. This demonstrates that, although our method is formulated on a vertex basis, it can still robustly capture primitive-based interactions.

\begin{figure}[tb]
  \centering
  \includegraphics[width=\columnwidth]{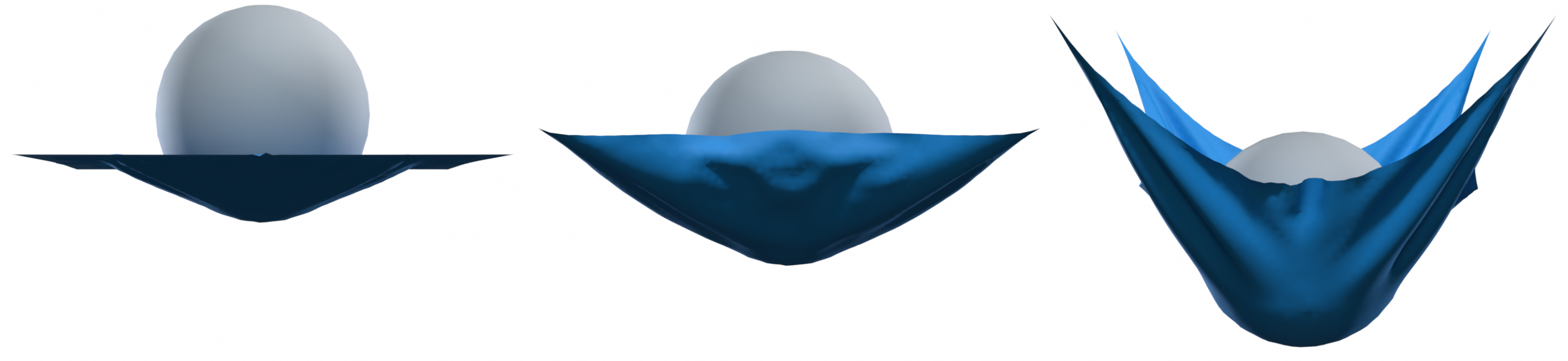}
  \caption{Collision between a heavy rigid steel sphere and a lightweight deformable cloth. Repulsive responses are activated before feasible-region projection, allowing contact forces to propagate through the cloth under a high mass-ratio setting.}
  \label{fig:drop-cloth}
\end{figure}

\noindent\textbf{Physically Active Test.}
We simulate collisions between a heavy rigid steel sphere and a lightweight deformable cloth to illustrate a high mass-ratio case. In this case, repulsive responses within the cloth are activated before vertices reach the boundary of the collision feasible region, allowing physically meaningful contact forces to propagate from the rigid sphere to the cloth before boundary projection (see Figure \ref{fig:drop-cloth}).

\noindent\textbf{Twisting Cloth.}
In this experiment, a single cloth sheet is subjected to continuous twisting, resulting in sharp folds and densely constrained regions that are challenging for collision handling (see Figure \ref{fig:cloth-twisting}).
Such configurations are known to be difficult for penalty-based solvers due to severe local vertex crowding, which often leads to secondary collisions.
The CCFR method maintains the prescribed offset distance throughout the twisting process. 


%

\begin{table*}[tb]
    \centering
    \caption{Performance results. Times are measured in seconds. 
    \emph{XPBD} measures the cost of constraint enforcement, including physics-based contact responses such as repulsion and friction.
    \emph{Broad Phase} includes both the construction and traversal of the AABB tree for collision detection.
    \emph{Narrow Phase} reports the cost of evaluating local convex constraints. 
    \emph{Resolution} corresponds to resolving infeasible configurations using our CCFR.} 
    \label{tab:performance}
    \resizebox{\textwidth}{!}{%
    \begin{tabular}{lrrrrrrrr}
        \toprule
        & \multicolumn{1}{c}{Number of} & \multicolumn{2}{c}{Frame Time} & \multicolumn{4}{c}{Breakdown of Average Frame Time}\\ 
        \cmidrule(lr){2-2} \cmidrule(lr){3-4} \cmidrule(lr){5-8}
        & Total Vert.  & Avrg. & Max. & XPBD & Broad Phase & Narrow Phase & Resolution \\
        \midrule      


        
        Twisting cloth (Figure \ref{fig:cloth-twisting}) & 15,908  & \num{0.902338} & \num{3.069347} & \num{0.844467} & \num{0.872114} & \num{0.250559} & \num{0.024683}  \\




        Soft Wires (Figure \ref{fig:strings}) & 40,000 & \num{0.391849} & \num{0.555964} & \num{0.017398} & \num{0.149785} & \num{0.180273} & \num{0.022698} \\

        Hair (Figure \ref{fig:hair}) & 33,095 & \num{6.639455} & \num{7.726633} & \num{1.691576} & \num{2.544423} & \num{0.471595} & \num{0.471512} \\ 

        Particles (Figure \ref{fig:particles}) & 3,182 & \num{0.026110} & \num{0.064368} & \num{0.008049} & \num{0.006509} & \num{0.007527} & \num{0.002347} \\ 

        Fully Coupled Codimensional (Figure \ref{fig:codimensional-all}) & 13,586 & \num{0.208647} & \num{0.305605} & \num{0.079522} & \num{0.026361} & \num{0.081292} & \num{0.016530} \\ 
        \bottomrule
    \end{tabular}%
    }
\end{table*}

\begin{figure*}[htb]
  \centering
  \includegraphics[width=\textwidth]{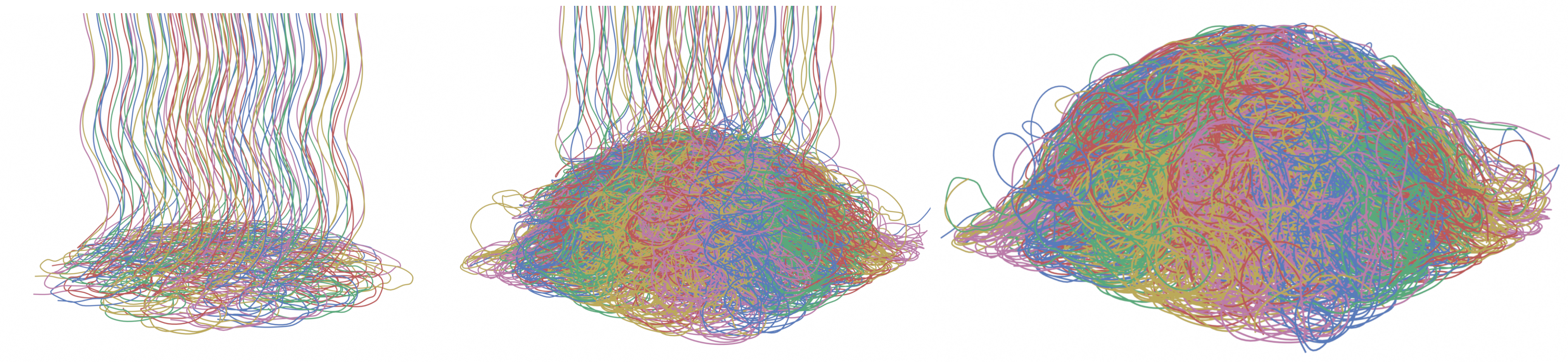}
  \caption{Wire-like simulation with 100 thin strings undergoing dense string--string, string--wall, and self-collisions.
    }
  \label{fig:strings}
\end{figure*}

\noindent\textbf{Soft Wires.}
We further evaluate the CCFR method on wire-like simulations consisting of 100 thin strings dropped into an enclosed environment with surrounding boundary walls (see Figures \ref{fig:teaser} and~\ref{fig:strings}).
This scenario involves frequent string--string and string--wall collisions, as well as extensive self-collision within each string.
%

\begin{figure*}[tb]
  \centering
  \includegraphics[width=\textwidth]{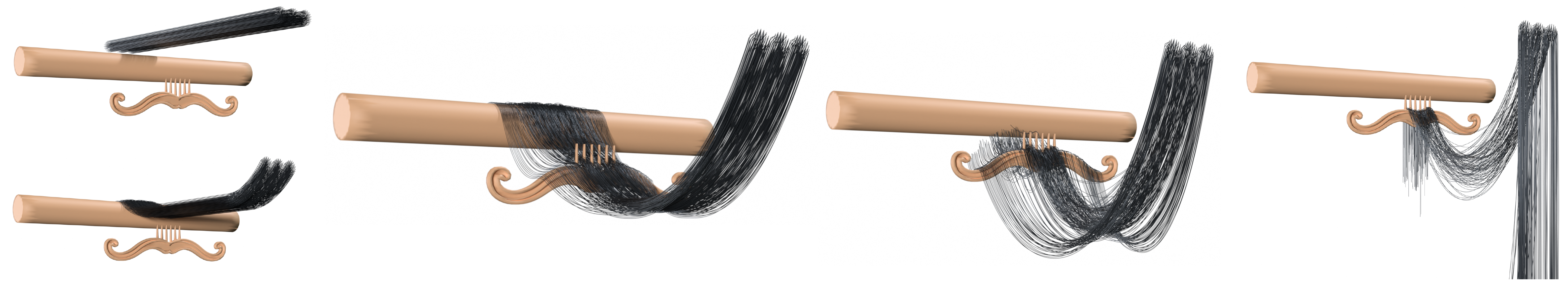}
  \caption{Hair collision scenario involving a bundle of 300 strands falling under gravity onto cylindrical and comb-shaped obstacles.
  From left to right: the initial configuration; collision with the cylindrical obstacle; interaction with the comb, where hair strands collide, entangle, and become locally trapped between the teeth; and the final state, in which strands caught by the comb remain clustered while those in free regions continue to fall.}
  \label{fig:hair}
\end{figure*}

\noindent\textbf{Hairs.}
We simulate a bundle of 300 hair strands interacting with a cylindrical obstacle (see Figure \ref{fig:hair}). 
The hairs fall under gravity and collide with the triangle mesh surfaces of the cylinder and the comb, inducing dense self-collisions and contact with the rigid surface.

\noindent\textbf{Particles.}
We demonstrate a particle simulation of 300 particles dropping onto a sphere (see Figure \ref{fig:particles}).
The particles are simulated as point primitives without intrinsic volume, while point-point interactions maintain regular offsets while preventing interpenetration.

\begin{figure*}[htb]
  \centering
  \includegraphics[width=\textwidth]{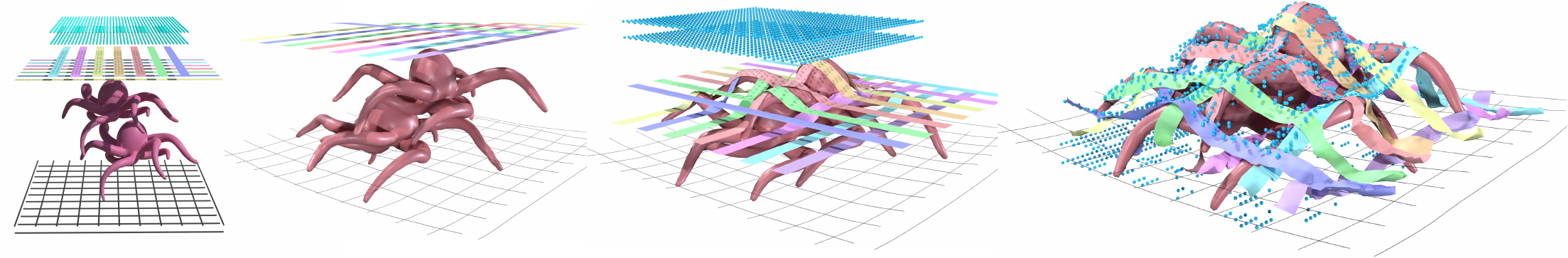}
  \caption{Coupled codimensional interactions involving a wire-like rod structure, deformable elastic octopuses, paper strips, and a column of particles. From left to right: the initial configuration; interaction between the wire structure and the elastic octopuses, where tentacles pass through and collide within the gaps between the strings; the paper strips draping over both the wire structure and the octopuses; and the final state, where a column of particles falls onto and interacts with the surface of the paper strips.}
  \label{fig:codimensional-all}
\end{figure*}

\noindent\textbf{Fully Coupled Codimensional Scenario.}
Finally, we test a fully coupled codimensional scenario involving multiple geometric representations in a single simulation (see Figures \ref{fig:teaser} and \ref{fig:codimensional-all}). 
A wire-like rod structure is placed at the bottom, above which two deformable octopuses are dropped, followed by a paper layer and finally a horizontal layer of particles.
%
This experiment demonstrates unified collision handling across strands, surfaces, volumes, and particles within a single codimensional scene, while maintaining non-interpenetrating configurations throughout the simulation.

\subsection{Performance and Behavioral Analysis}
\noindent\textbf{Performance.}
Table \ref{tab:performance} reports the performance of the proposed CCFR method across a range of collision scenarios.
All timings are measured per frame and averaged over the full simulation.
Overall, the total frame time is dominated by collision detection, with the broad and narrow phases accounting for most of the computational cost in the majority of scenes.
In contrast, the cost of the proposed \emph{Resolution} stage remains consistently small, even in large-scale or densely interacting scenarios.
This behavior follows from resolving collisions through lightweight local projections onto convex feasible regions, without iterative nonlinear optimization.
%

These results indicate that the proposed collision formulation adds minimal overhead relative to existing constraint enforcement and collision detection steps, while remaining suitable for efficient and large-scale elastic simulations.

\begin{figure}[tb]
  \centering
  \includegraphics[width=\columnwidth]{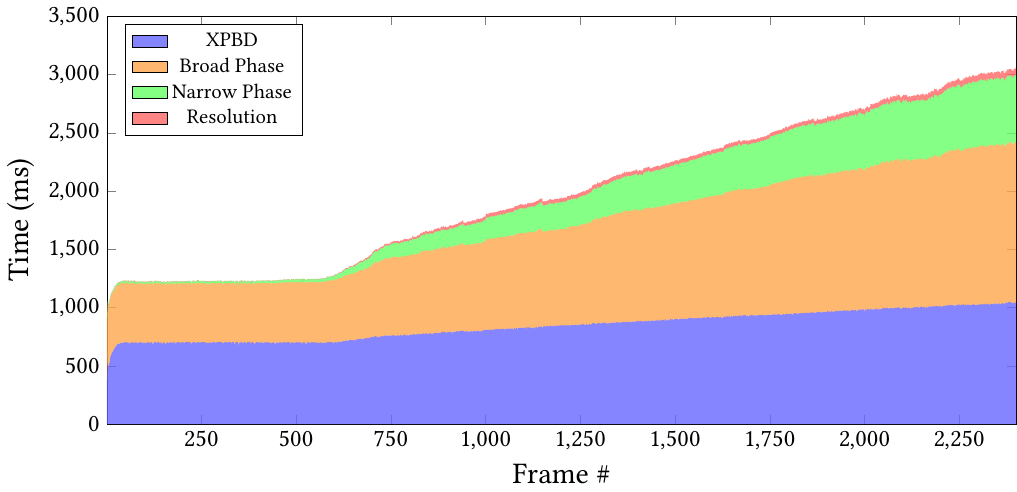}
  \caption{Timing breakdown per frame for Figure \ref{fig:cloth-twisting} in Table \ref{tab:performance}. The horizontal axis shows the frame index, while the vertical axis reports the accumulated runtime of each component, stacked from bottom to top as \textit{XPBD}, \textit{Broad Phase}, \textit{Narrow Phase}, and \textit{Resolution}.
  The visualization shows that the proposed projection-based \textit{Resolution} stage contributes only a small portion of the total runtime throughout the sequence.}
  \label{fig:timing-breakdown}
\end{figure}

\noindent\textbf{Detailed Performance Breakdown.}
The performance table in the previous paragraph shows that several test cases exhibit
a noticeable discrepancy between their Avg. and Max. runtime per frame.
To better understand the origin of this behavior, we examine the ``Twisting cloth''
sequence (Figure~\ref{fig:cloth-twisting}). 
Figure \ref{fig:timing-breakdown} visualizes the runtime decomposition of this sequence,
where the total frame time is stacked from bottom to top as
\textit{XPBD}, \textit{Broad Phase}, \textit{Narrow Phase}, and \textit{Resolution}.
%
In the most extreme frames, the number of collision candidates reaches up to $8.3 \times 10^6 $, leading to a substantial increase in detection cost.
This explains the large gap between the Avg. and Max. runtimes reported in Table~\ref{tab:performance}. Although the early frames, before the cloth begins to twist tightly, are processed efficiently, the later frames experience higher runtimes as the number of active contact candidates gradually increases.

Overall, this breakdown shows the worst-case behavior observed in the
experiment is not inherent to the proposed projection-based solver in the
\textit{Resolution} stage, but is dominated by contact detection costs, suggesting further optimization of broad and narrow phase collision detection, and candidate
pruning is the most effective direction for improving robustness against
pathological frames.

\begin{figure}[tb]
  \centering
  \includegraphics[width=\columnwidth]{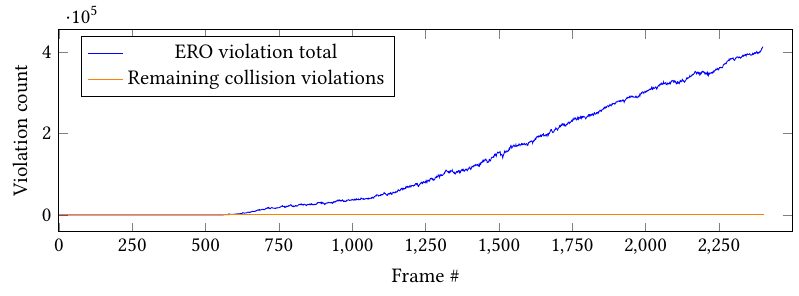}
  \caption{Per-frame statistics of collision violations during the XPBD simulation. The blue curve shows the total number of detected violations during constraint processing, while the orange curve shows the number of remaining violations after all collision projections have been completed. The final violation count consistently returns to zero, demonstrating robust non-penetration enforcement.}
  \label{fig:num-violation}
\end{figure}

\noindent\textbf{Non-Penetration Guarantee Under Dense Contact.}
Figure \ref{fig:num-violation} reports the total number of collision violations detected during the XPBD simulation, together with the number of remaining violations after all collision handling steps have been completed for every vertex. The results demonstrate that the non-penetration condition is consistently satisfied at the end of each frame.
As the cloth undergoes increasing twisting and becomes more densely packed, a larger number of vertices are displaced outside the feasible convex region due to the influence of other constraint projections. Nevertheless, after applying the Plesn\'ik projection, all remaining violations are resolved, confirming that no remaining violations were observed at the end of each frame in our experiments.

\begin{figure}[tb]
  \centering
  \includegraphics[width=\columnwidth]{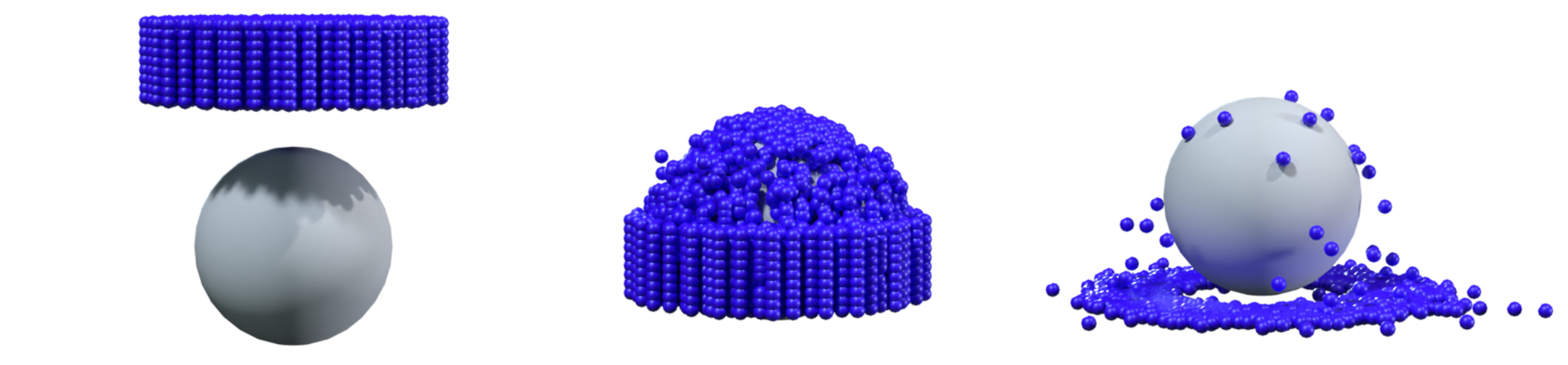}
  \caption{Particle collision simulation. Particles are simulated as points, while spheres are used only for visualization.}
  \label{fig:particles}
\end{figure}

\begin{figure}[tb]
  \centering
  \includegraphics[width=\columnwidth]{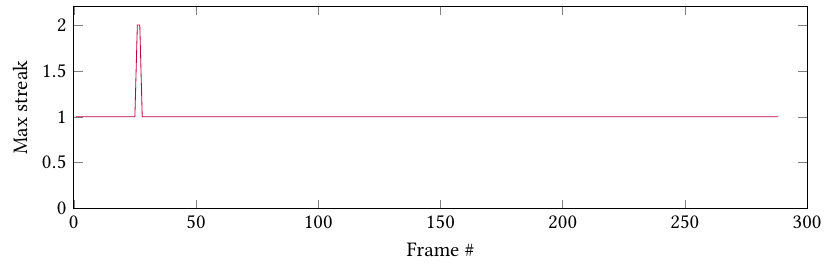}
  \caption{Number of consecutive halting events per frame for the particle configuration shown in Figure~\ref{fig:particles}.
While a short sequence of two consecutive halts is observed, the simulation continues to progress afterward without persistent halting, even from a highly compressed initial state.
}
  \label{fig:num-halting}
\end{figure}

\noindent\textbf{Temporal Behavior of Halting Events.}
Figure \ref{fig:num-halting} plots the number of consecutive halting events for each frame in the example shown in Figure~\ref{fig:particles}. Although two consecutive halting events are observed at some point in the simulation, the number of collision points involved remained limited, accounting for at most $3 \%$ of the total collision points. The system subsequently progresses without sustained halting, despite starting from a densely accumulated particle configuration.

\subsection{Comparison with Other Methods}
\begin{figure*}[htb]
  \centering
  \includegraphics[width=\textwidth]{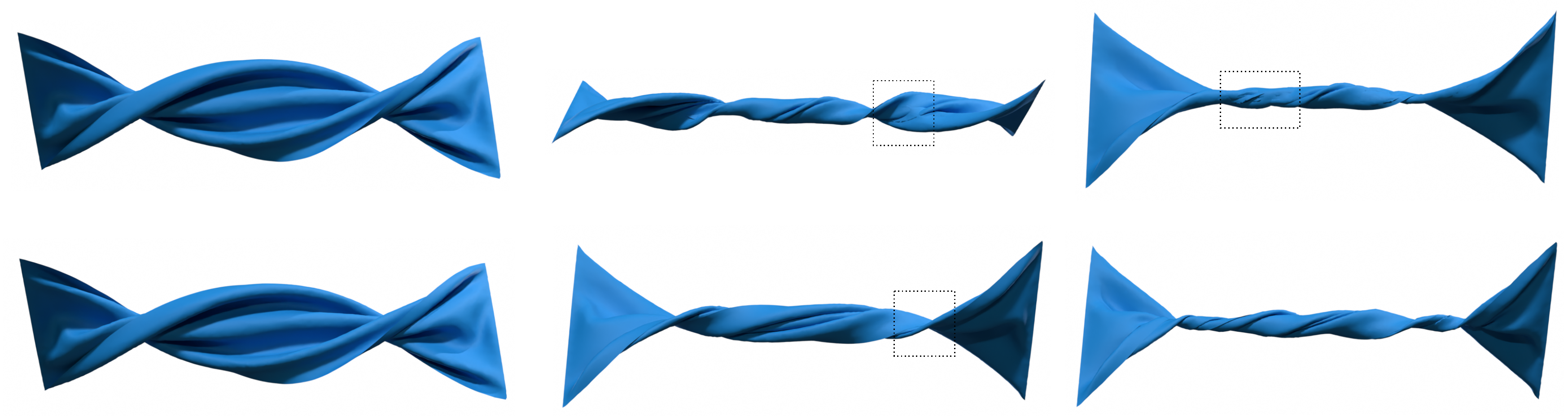}
  \caption{Twisting deformation of a single cloth sheet.
(Top) Conventional post-process collision handling in XPBD.
As the deformation progresses, repeated secondary collisions gradually introduce local intersections in densely folded regions (highlighted by dashed boxes).
(Bottom) The proposed CCFR method maintains non-penetrating configurations throughout the twisting sequence.    }
  \label{fig:cloth-twisting}
\end{figure*}

\noindent\textbf{Comparison with post-process method.}
We compare our method with a conventional post-process-based collision handling approach in XPBD.
While the post-process method can resolve simple self-collisions, it gradually loses consistency once dense contacts and repeated secondary collisions accumulate during large twisting deformations.
As shown in Figure \ref{fig:cloth-twisting}, local intersections and collision artifacts increasingly appear as the deformation progresses.
In contrast, our method maintains collision-free configurations throughout the simulation, enabling sustained twisting deformation.

\begin{figure}
  \centering
  \includegraphics[width=.5\columnwidth]{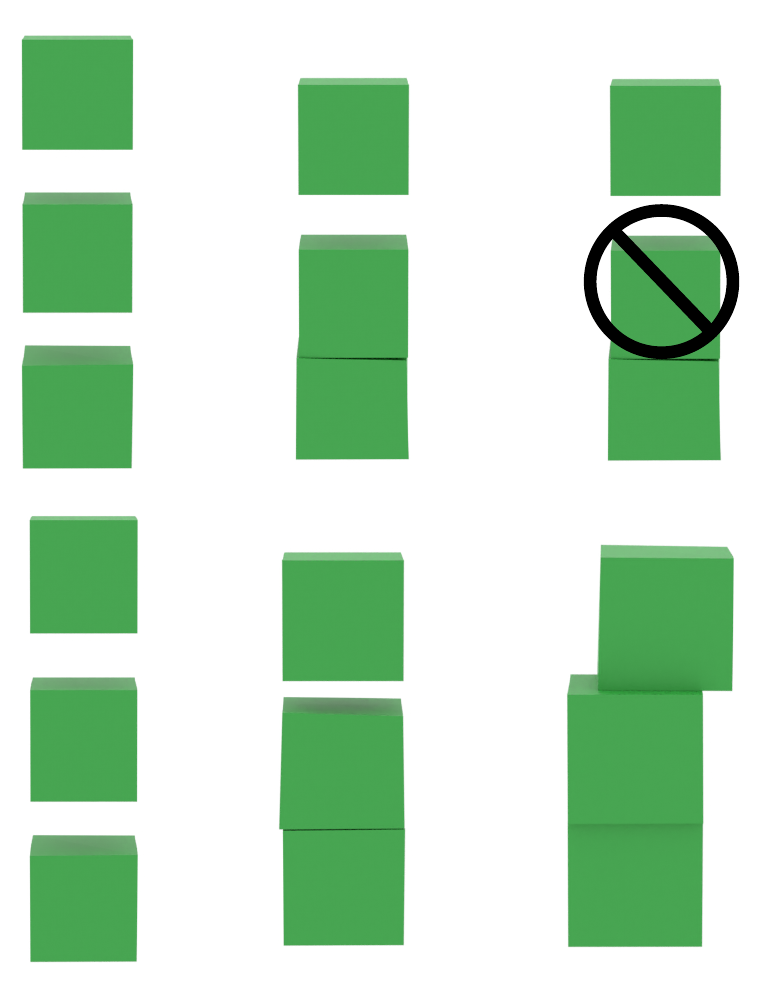}
  \caption{Comparison with Lu et al. Frames 48, 60, and 140 are shown from left to right. (Top) In the method of Lu et al., the minimum TOI becomes zero after frame 65, causing the simulation to stop proceeding. (Bottom) CCFR continues the simulation without interruption.}
  \label{fig:comp-lu}
\end{figure}

\noindent\textbf{Comparison with XPBD-based IPC.}
We compare CCFR with the XPBD-based IPC method of Lu et al.~\shortcite{Lu23} on collisions between three tetrahedral cubes, as demonstrated in Figure \ref{fig:comp-lu}.
Their method maintains non-penetration through ACCD-based TOI interpolation after both predicted position calculation and XPBD constraint updates.
However, unlike the braiding scenarios considered in their work, falling cases with persistent gravity-induced contact may cause the minimum TOI to remain zero, preventing further positional updates.
In contrast, CCFR continues to advance under sustained contact without ACCD-based TOI interpolation, while achieving lower runtime cost, as shown in Table~\ref{tab:comp-runtime}.

\begin{table}
  \centering
  \caption{Comparison with Lu et al. \cite{Lu23}: The runtime (in seconds) is reported as the average per frame. The \emph{Collision Handling} includes both narrow phase detection and resolution. }
  \label{tab:comp-runtime}
  \resizebox{.9\columnwidth}{!}{%
  \begin{tabular}{@{}lrrrr@{}}
    \toprule
    & Frame Time & Additive CCD & Collision Handling \\
    \midrule
    Lu et al.'s & \num{0.014399} & \num{0.009734}&  \num{0.003526} \\
    CCFR (Ours) & \num{0.003792} & --- &  \num{0.002512} \\
    \bottomrule
  \end{tabular}
  }
\end{table}

\subsection{Limitations}
The proposed method is more conservative than CCD-based approaches, potentially causing reduced feasible regions and locally inconsistent motion in dense or fast-moving scenarios.
The formulation also employs a fixed displacement bound per step, which may limit attainable velocities under high acceleration.
Numerical robustness under single-precision arithmetic and integration with high-stiffness formulations requiring second-order information, such as Hessian-based methods, remain open challenges.

\section{Conclusion}
We have presented \emph{Convex Collision-Free Regions} (CCFR), a collision formulation that enforces non-penetration through projection onto local convex feasible regions.
The method enables geometric collision avoidance without relying on nonlinear optimization
Experiments demonstrate effective collision handling on cloth, hair, yarn, and particle systems.
By separating geometric feasibility from physical contact response, CCFR handles non-penetration through projection, while repulsion and friction are treated independently within the dynamics solver.
This avoids penetration-depth estimation, barrier parameter tuning, and iterative nonlinear solves, resulting in a lightweight projection-based collision formulation.
Overall, CCFR provides an explicit geometric feasibility framework for low-overhead non-penetrating simulation.

\bibliographystyle{ACM-Reference-Format}
\bibliography{thesis}

\clearpage


\begin{figure}[tb]
  \centering
  \includegraphics[width=.6\columnwidth]{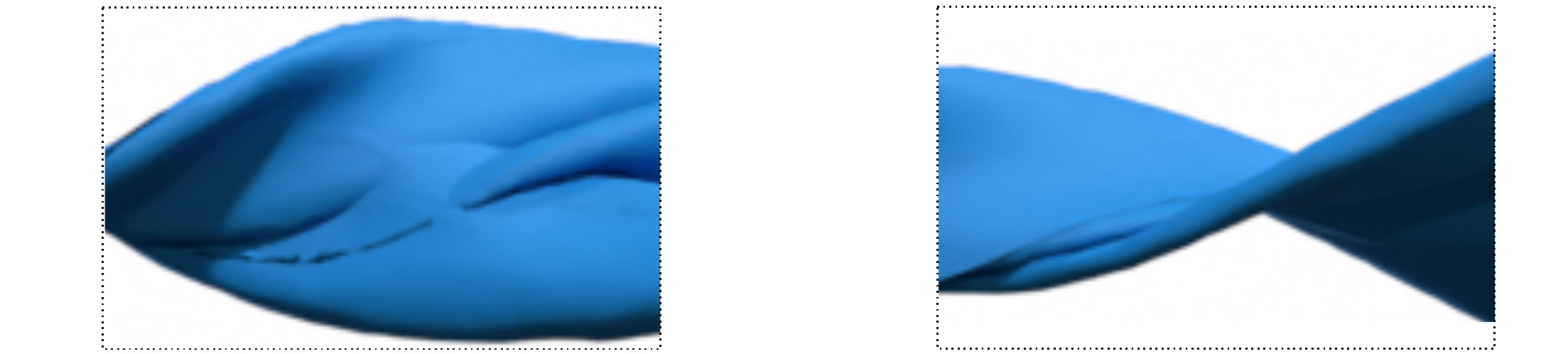}
  \caption{Close-up view of the dashed regions highlighted in the middle column of Fig.~\ref{fig:cloth-twisting}.
(Left) Conventional post-process collision handling in XPBD.
As secondary collisions accumulate, local intersections and collision artifacts appear in densely folded regions.
(Right) The proposed CCFR method maintains non-penetrating configurations under the same deformation.}
  \label{fig:comp-twisting}
\end{figure}

\clearpage
\appendix

\section{Orthogonal Projection Algorithm}
\label{appendix-sec:orthogonal}

This section provides an overview of Plesn\'{i}k's algorithm.
Although originally developed for finding the orthogonal projection onto the solution set of a system of linear equations, the algorithm is employed here to project vertices onto the boundary of the convex feasible region described in Section 3.2 in the paper. 
The discussion begins with the fundamental principles of the algorithm, followed by its pseudocode.

\begin{figure}
  \centering
  \includegraphics[width=\columnwidth]{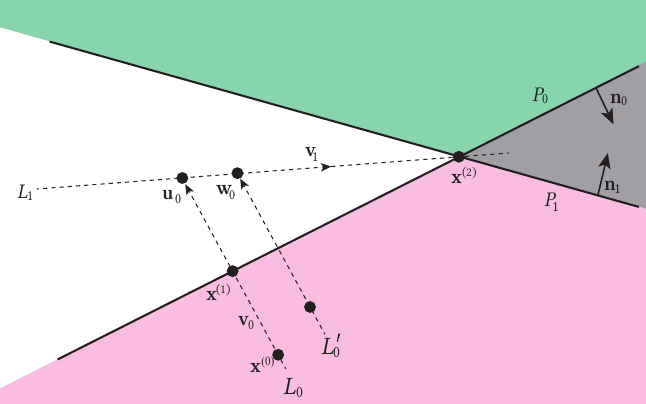}
  \caption{Plesn\'ik's algorithm for two constraints. 
    Starting from the initial point $\mathbf{x}^{(0)}$, the first projection $\mathbf{x}^{(1)}$ is computed onto the boundary hyperplane $P_0$, which defines the pink feasible half-space.
    The second hyperplane $P_1$ defines the green feasible half-space to be incorporated.
    For the $m=0$ step, the direction vector $\mathbf{v}_0$ coincides with the normal vector $-\mathbf{n}_0$ of $P_0$.
    At the $m=1$ step, the ERO line $L_1$ is constructed to incorporate the new boundary $P_1$, and the intersection $\mathbf{x}^{(2)} = L_1 \cap P_0$ yields the orthogonal projection of $\mathbf{x}^{(0)}$ onto the updated feasible region $S_1 = P_0 \cap P_1$.}
  \label{fig:orthogonal-algorithm}
\end{figure}

\subsection{The algorithm proposed by Plesn\'ik}
The goal of this algorithm is to recursively compute the orthogonal projection $\mathbf{x}^{(m+1)}$ of a given point $\mathbf{x}^{(0)}$ onto the solution set $S = P_0 \cap \cdots \cap P_{m}$ of a system of linear equations, where each $P_i$ denotes a hyperplane.
An equiresidual orthogonal (ERO) line is defined as a line along which the absolute values of the residuals with respect to a given set of equations are equal for any point on the line.
The algorithm relies on this concept, in particular on the property that an ERO line passes through the solution set $S$ and is orthogonal to~$S$.

Let us consider the recursive procedure from step~$m-1$ to step~$m$ ($m \ge 1$).
We assume that the direction vector $\mathbf{v}_{m-1}$ of the ERO line $L_{m-1}$ for the first $m$ linearly independent equations has already been determined. 
Next, we determine two equiresidual points $\mathbf{u}_{m-1}$ and $\mathbf{w}_{m-1}$.
The point $\mathbf{u}_{m-1}$ lies on the existing ERO line $L_{m-1}$ and satisfies equal residuals with respect to $P_0$ and the newly added hyperplane $P_m$.
The point $\mathbf{w}_{m-1}$ is constructed by recursively enforcing the equiresidual condition through a sequence of parallel ERO lines, starting from $\mathbf{w}_0 = \mathbf{x}_0 + \mathbf{n}_m$.

If $\mathbf{u}_{m-1} = \mathbf{w}_{m-1}$ and $\mathbf{x}^{(m+1)} \in P_m$, then the equation defining $P_m$ is linearly dependent on the previous equations and is therefore skipped.
If $\mathbf{u}_{m-1} = \mathbf{w}_{m-1}$ and $\mathbf{x}^{(m+1)} \notin P_m$, the system is inconsistent and the algorithm terminates; however, in the present setting, we assume that no inconsistent constraints are present.
If $\mathbf{u}_{m-1} \neq \mathbf{w}_{m-1}$, the vectors $\mathbf{n}_0, \ldots, \mathbf{n}_{m}$ are linearly independent, and a new direction vector $\mathbf{v}_m = \mathbf{w}_{m-1} - \mathbf{u}_{m-1}$ is defined.
The corresponding ERO line is given by $L_m = \{\mathbf{u}_{m-1} + \lambda \mathbf{v}_m \mid \lambda \in \mathbb{R}\}$.
The intersection $\mathbf{x}^{(m+1)} = L_m \cap P_0$ then provides the orthogonal projection of $\mathbf{x}^{(0)}$ onto the updated solution set $S_m$.
By repeating this procedure, the solution is constructed incrementally by computing orthogonal projections onto intersections of successively augmented sets of hyperplanes, as illustrated in Figure~\ref{fig:orthogonal-algorithm}.

\subsection{Algorithm}

Algorithm \ref{alg:orthogonalProjection} performs the core projection computation using the ERO line method when a new constraint is added to the system of equations.
Lines 1--6 process the first violated constraint encountered and compute a direct orthogonal projection of $\mathbf{x}$ onto the corresponding constraint plane.
Line 8 sets the current number of previously identified independent constraints.
Lines 9--15 compute the equiresidual point $\mathbf{u}$ on the existing ERO line based on the current position.
The original steps for handling the zero-denominator case---designed to distinguish between dependent and inconsistent constraints via sign flipping and the computation of $\mathbf{u}$ and $\mathbf{w}$---are not employed, as the CCFR method assumes consistent inputs and does not require such classification.
Therefore, when the denominator is close to zero, the constraint is directly treated as redundant and skipped (line 14).
Line 16 initializes the auxiliary starting point $\mathbf{w}$.
The loop in lines 17--25 recursively applies the previously stored ERO line direction vectors to update $\mathbf{w}$, thereby determining the equiresidual point $\mathbf{w}$ for the new constraint.
Line~26 computes the direction vector of the new ERO line.
If $\|\mathbf{v}_{\mathrm{new}}\|$ is non-zero, linear independence of the constraints is confirmed (line 30).
Lines~27--36 compute the intersection of the new ERO line $L_m$ with the first constraint plane $P_0$.
This intersection point corresponds to the new orthogonal projection onto the intersection of all constraints processed so far.
Line 35 stores the updated direction vector and constraint data for use in the subsequent iteration.

\begin{algorithm}[!t]
\setlength{\algomargin}{0.8em}
\SetAlgoNoLine
\LinesNumbered
\caption{orthogonalProjection (per constraint associated with the vertex)}
\label{alg:orthogonalProjection}
\KwIn {
$\tilde{\mathbf{x}}$: predicted position of the vertex, 
$\mathbf{x}^{(0)}$: initial position of the vertex, 
$\mathbf{n}$: the normal vector of the new constraint, 
$b$: RHS of the new constraint, 
$\mathcal{V, A, B}$: lists of previous ERO directions and constraints,
$\varepsilon$:~tolerance (e.g., $10^{-12}$).
}
\KwOut {$\tilde{\mathbf{x}}$: updated to the orthogonal projection.}
\vspace{0.5em}
\hrule
\vspace{0.5em}
\eIf{$\mathcal{A}$ is empty}{
    $\mathbf{v}_0 \gets \mathbf{n}$\;
    $\beta_0 \gets \mathbf{n} \cdot \mathbf{v}_0$\;
    \textbf{if} {$\|\beta_0\| < \varepsilon$ \textbf{then return} $\tilde{\mathbf{x}}$\;}
    
    $\tilde{\mathbf{x}} \gets \tilde{\mathbf{x}} + \dfrac{b - \mathbf{n} \cdot \tilde{\mathbf{x}}}{\beta_0} \, \mathbf{v}_0$\;
    
    $\mathcal{V} \gets \mathcal{V} \cup \{\mathbf{v}_0\}; \quad \mathcal{A} \gets \mathcal{A} \cup \{\mathbf{n}\}; \quad \mathcal{B} \gets \mathcal{B} \cup \{b\}$\;
}
{
    $m \gets |\mathcal{V}|$\;
    $\mathbf{n}_0 \gets \mathcal{A}[0]$; \quad $b_0 \gets \mathcal{B}[0]$\;
    $\mathbf{n}_m \gets \mathbf{n}$; \quad $b_m \gets b$\;
    $\Delta \mathbf{n} \gets \mathbf{n}_m - \mathbf{n}_0$\;
    
    $\beta_\mathbf{u} \gets \Delta \mathbf{n} \cdot \mathcal{V}[{m-1}]$\;
    $\alpha_\mathbf{u} \gets b_m - \mathbf{n}_m \cdot \tilde{\mathbf{x}}$\;
    
    \textbf{if} {$\|\beta_\mathbf{u}\| < \varepsilon$ \textbf{then return} $\tilde{\mathbf{x}}$\;}
    
    $\mathbf{u} \gets \tilde{\mathbf{x}} + \dfrac{\alpha_\mathbf{u}}{\beta_\mathbf{u}} \, \mathcal{V}[{m-1}]$\;
    $\mathbf{w} \gets \mathbf{x}^{(0)} + \mathbf{n}_m$\;
    
    \For{$j \gets 1$ \KwTo $m$} {
        $\mathbf{n}_{j} \gets \mathcal{A}[j]; \quad b_{j} \gets \mathcal{B}[j]$\;
        $\Delta \mathbf{n}_j \gets \mathbf{n}_{j} - \mathbf{n}_0$\;
        $\beta_\mathbf{w} \gets \Delta \mathbf{n}_j \cdot \mathcal{V}[{j-1}]$\;
        \If{$\|\beta_\mathbf{w}\| > \varepsilon$} {
            $\alpha_\mathbf{w} \gets b_{j} - b_0 - \Delta \mathbf{n}_j \cdot \mathbf{w}$\;
            $\mathbf{w} \gets \mathbf{w} + \dfrac{\alpha_\mathbf{w}}{\beta_\mathbf{w}} \,\mathcal{V}[{j-1}]$\;
        }
    }
    
    $\mathbf{v}_\mathrm{new} \gets \mathbf{w} - \mathbf{u}$\;
    \If{$\|\mathbf{v}_\mathrm{new}\| > \varepsilon$} {
        $\beta_\mathrm{new} \gets \mathbf{n}_0 \cdot \mathbf{v}_\mathrm{new}$\;
        \eIf{$\|\beta_\mathrm{new}\| < \varepsilon$} {
            $\tilde{\mathbf{x}} \gets \mathbf{u}$\;
        }
        {
            $\alpha_\mathrm{new} \gets b_0 - \mathbf{n}_0 \cdot \mathbf{u}$\;
            $\tilde{\mathbf{x}} \gets \mathbf{u} + \dfrac{\alpha_\mathrm{new}}{\beta_\mathrm{new}} \, \mathbf{v}_\text{new}$\;
        }
        $\mathcal{V} \gets \mathcal{V} \cup \{\mathbf{v}_\mathrm{new}\}; \quad \mathcal{A} \gets \mathcal{A} \cup \{\mathbf{n}_k\}; \quad \mathcal{B} \gets \mathcal{B} \cup \{b_k\}$\;
    }
}
\Return{$\tilde{\mathbf{x}}$}\;
\end{algorithm}

\section{Collision Handling Algorithm}
This section presents the core collision handling algorithm of the proposed CCFR method.
Building upon a standard broad phase and narrow phase collision detection
pipeline, we detail the proposed contact response algorithm that enforces
non-penetration via convex feasible regions.

First, the broad phase is executed to coarsely identify potentially
colliding primitive pairs, using a bounding volume hierarchy.
Subsequently, the narrow phase identifies primitive pairs whose closest points lie within the vertex-centered search neighborhoods of radius~$r$, and extracts the associated convex boundary constraints $c \in \mathcal{C}$.
The feasible region is constructed implicitly by this set of constraints, without explicitly computing their intersections, thereby retaining only the proximity-related quantities: the feasible constraint values $d_c$, the contact normal $\mathbf{n}_c$, and the geometric separation measure $\phi_c$.
The algorithmic details of the narrow phase are provided in Algorithm~\ref{alg:NarrowPhase_detail}.

After external forces and internal energy minimization are applied, vertices may temporarily leave their respective feasible regions.
Such vertices are regarded as collision violations and are returned to feasibility by projecting onto the corresponding active constraints.
This procedure is managed by Algorithm \ref{alg:projectOntoConvexRegion}, which invokes Plesn\'{i}k's orthogonal projection routine for the core calculation.

Algorithm \ref{alg:projectOntoConvexRegion} orchestrates the outer iterative process that computes the projected position $\mathbf{x}^*$ from the updated position $\mathbf{x}'$ onto the feasible convex region.
Internally, the algorithm maintains a predicted position
$\mathbf{x}_{\mathrm{pred}}$, which is iteratively refined until convergence.
The process iterates up to $n_{\mathrm{maxIterations}}$ (line \ref{line:convex-iteration-start}) to ensure convergence when multiple constraints are simultaneously active.
Notably, the constraints are pre-sorted by their violation depth in ascending order (lines~\ref{line:convex-initialize-start}--\ref{line:convex-initialize-end}); by addressing the most deeply violated constraints first, the iterative process tends to converge rapidly. 
Consequently, setting $n_{\mathrm{maxIterations}} = 1$ is often sufficient for most practical applications.
The predicted position $\tilde{\mathbf{x}}$ and the initial position for the iteration, $\mathbf{x}^{(0)}$, are initialized (lines \ref{line:convex-set-x}--\ref{line:convex-set-x0}).
Subsequently, the sets for the equiresidual orthogonal (ERO) line direction vectors $\mathcal{V}$, normals $\mathcal{A}$, and right-hand sides $\mathcal{B}$ are initialized (line~\ref{line:convex-set-vab}).
Line \ref{line:convex-re-evaluate} re-evaluates the constraint at the predicted solution. 
Constraints that are already satisfied ($\rho_k \ge 0.0$) are bypassed (line~\ref{line:convex-not-violated}).
Lines~\ref{line:convex-violated-set-start}--\ref{line:convex-violated-set-end} configure the parameters for the violated constraint, and line~\ref{line:convx-call-orthogonal} invokes the subroutine $\mathrm{orthogonalProjection}$.
Lines~\ref{line:convex-safety-start}--\ref{line:convex-safety-end} enforce a numerical safety margin as discussed in Section 3.2 in the paper. 
The loop terminates if no constraints are violated (line~\ref{line:convex-no-violation}), indicating successful convergence to the feasible region.
If the maximum number of iterations is reached without convergence (line~\ref{line:convex-reach-max-start}), the position reverts to the current valid state.

\begin{algorithm}[!t] 
\SetAlgoNoLine
\LinesNumbered
\caption{projectOntoConvexRegion (per vertex)}
\label{alg:projectOntoConvexRegion}
\KwIn {
$\mathbf{x}'$: updated position of the vertex,
$\mathbf{x}$: current position of the vertex,
$\mathcal{C}$: a set of convex boundary constraints associated with this vertex, 
$\varepsilon$: tolerance (e.g., $10^{-6}$).}
\KwOut {$\mathbf{x}'$: updated configuration.}
\vspace{0.5em}
\hrule
\vspace{0.5em}
\tcp{Each constraint $c_k \in \mathcal{C}$ stores $(d_k,\mathbf{n}_k,\phi_k)$.}
\textbf{if} {$|\mathcal{C}| = 0$ \textbf{then return} $\mathbf{x}'$}\tcp*{No constraints exist.}
Compute $\rho_k = (\mathbf{x}'-\mathbf{x})\!\cdot\!\mathbf{n}_k + d_k$ for all $c_k\!\in\!\mathcal{C}$\;\label{line:convex-initialize-start}
Sort $\mathcal{C}$ by $\rho_k$ in ascending order\; \label{line:convex-initialize-end}
\For{$n = 1$ \KwTo $n_\mathrm{maxIterations}$}{\label{line:convex-iteration-start}
    
    $\tilde{\mathbf{x}} \gets \mathbf{x}'$; \quad $n_\mathrm{violated} \gets 0 $\;\label{line:convex-set-x}
    $\mathbf{x}^{(0)} \gets \mathbf{x}'$\tcp*{Store initial position.}\label{line:convex-set-x0}
    
    $\mathcal{V}, \mathcal{A}, \mathcal{B} \gets \emptyset$\;\label{line:convex-set-vab}

    \ForEach{$(d_k, \mathbf{n}_k) \in \mathcal{C}$}{
        $\rho_k \gets (\tilde{\mathbf{x}} - \mathbf{x}) \cdot \mathbf{n}_k + d_k$\tcp*{Re-evaluate constraint.}\label{line:convex-re-evaluate}
        \textbf{if} {$\rho_k \ge 0.0$ \textbf{then} continue}\tcp*{Not violated.}\label{line:convex-not-violated}
        
        $b \gets \mathbf{n}_k \cdot \mathbf{x} - d_k$\;\label{line:convex-violated-set-start}
        $n_\mathrm{violated} \gets n_\mathrm{violated} + 1$\;\label{line:convex-violated-set-end}    
        
        orthogonalProjection($\tilde{\mathbf{x}}, \mathbf{x}^{(0)}, \mathbf{n}_k, b, \mathcal{V}, \mathcal{A}, \mathcal{B}$)\;\label{line:convx-call-orthogonal}
        \tcp{See Alg.~\ref{alg:orthogonalProjection}.}
    
        $\Delta \mathbf{x}_\mathrm{new} \gets \tilde{\mathbf{x}} - \mathbf{x}$\;
    
        \eIf{$\| \Delta \mathbf{x}_\mathrm{new} \| > \varepsilon$}{\label{line:convex-safety-start}
            $\tilde{\mathbf{x}} \gets \mathbf{x} + (\| \Delta \mathbf{x}_\mathrm{new} \| - \varepsilon) \dfrac{\Delta \mathbf{x}_\mathrm{new}}{\| \Delta \mathbf{x}_\mathrm{new} \|}$\;
        }
        {
            $\tilde{\mathbf{x}} \gets \mathbf{x}$\;
        }\label{line:convex-safety-end}
    }

    \textbf{if} {$n_\mathrm{violated} = 0$ \textbf{then return} $\mathbf{x}'$}\tcp*{Convergence achieved.}\label{line:convex-no-violation}
    $\mathbf{x}' \gets \tilde{\mathbf{x}}$\;
}

    $\mathbf{x}' \gets \mathbf{x}$\tcp*{Fallback for non-convergence.}\label{line:convex-reach-max-start}
\end{algorithm}

\section{Details of Narrow Phase Algorithm}
\label{appendix-sec:narrow-phase}
The details of the narrow phase algorithm are provided in Algorithm~\ref{alg:NarrowPhase_detail}.
For each primitive pair, the closest points are computed (lines 4, 18--19, and 35), and the contact normals $\mathbf{n}$ are determined (lines 5, 20, and 36).
Utilizing these normals, the geometric separation measures $\phi$ are evaluated (lines 6, 21, and 37).
The primary criterion for collision detection in this work is whether the closest points lie within a sphere centered at each vertex (lines 7, 22, and 38).
Specifically, a collision is considered imminent when the geometric separation $\phi$ falls below $2r$, representing the case where the protective shells of radius $r$ associated with each primitive begin to overlap.
It should be noted, however, that this radius provides no guarantee on the convergence of internal constraints, potentially necessitating a considerably smaller step size for convergence.
Primitive pairs identified within this sphere are stored as a set of linear constraints associated with the vertex.
Each constraint represents the distance between the closest points $d_k$, scaled by the inverse masses $w$ along the direction connecting these points (lines 8--10, 23--26, and 39--41).

\begin{algorithm}[!t]
\setlength{\algomargin}{0.8em}
\caption{Narrow Phase (per vertex)} 
\label{alg:NarrowPhase_detail} 
\SetAlgoNoLine
\LinesNumbered
\KwIn{
    $p$: active vertex with the current position $\mathbf{x}$, 
    $r$: query radius, $\xi$: offset,
    $\mathcal{T}$: passive face set,
    $e$: incident edges, $\mathcal{E}$: passive edge set,
    $f$: incident faces, $\mathcal{Q}$: passive vertex set. 
}
\KwOut{
$\mathcal{C}$: a set of convex boundary constraints associated with this vertex.
}
\vspace{0.5em}
\hrule
\vspace{0.5em} 
$\mathcal{C} \gets \emptyset$ 

\ForEach{passive face $f \in \mathcal{T}$}{ 
    \textbf{if} $p \subset f$ \textbf{then} continue\; 
    $\mathbf{x}_q \gets \mathrm{closestPointOnFace}(f, p)$\;
    $\mathbf{n} \gets \mathbf{x} - \mathbf{x}_q$\;
    $\phi(p, f) \gets \|\mathbf{n}\| - \xi$\;
    \If{$\phi(p, f) < 2r$}{
        $w_f \gets \sum_{v \in f}{(\mathrm{barycentric}_v w_v)}$\; 
        $\mathbf{n}_k \gets \mathbf{n} / \|\mathbf{n}\|$\; 
        $d_k \gets \frac{w_p}{w_p + w_f}\,\phi(p, f)$\; 
        $\phi_k \gets \phi(p, f)$\;
        $\mathcal{C} \gets \mathcal{C} \cup \{(d_k, \mathbf{n}_k, \phi_k)\}$\;
    }
}

\ForEach{active edge $e \in \mathrm{AdjEdges}(p)$}{
    \ForEach{passive edge $e' \in \mathcal{E}$}{
        \textbf{if} $e$ and $e'$ share a vertex \textbf{then} continue\;
        $\mathbf{x}_\mathrm{A} \gets \mathrm{closestPointOnEdge}(e, e')$\;
        $\mathbf{x}_\mathrm{B} \gets \mathrm{closestPointOnEdge}(e', e)$\;
        $\mathbf{n} \gets \mathbf{x}_\mathrm{A} - \mathbf{x}_\mathrm{B}$\;
        $\phi(e, e') \gets \|\mathbf{n}\| - \xi$\;
        \If{$\phi(e, e') < 2r$}{
            $w_e \gets \sum_{v \in e}{(\mathrm{barycentric}_v w_v)}$\;
            $w_{e'} \gets \sum_{v \in e'}{(\mathrm{barycentric}_v w_v)}$\;
            $\mathbf{n}_k \gets \mathbf{n} / \|\mathbf{n}\|$\; 
            $d_k \gets \frac{w_e}{w_e + w_{e'}}\,\phi(e, e')$\; 
            $\phi_k \gets \phi(e, e')$\;
            $\mathcal{C} \gets \mathcal{C} \cup \{(d_k, \mathbf{n}_k, \phi_k)\}$\;
        }
    }
}

\ForEach{active face $f_p \in \mathrm{AdjFaces}(p)$}{
    \ForEach{passive vertex $q \in \mathcal{Q}$}{
        \textbf{if} $q \subset f_p$ \textbf{then} continue\;
        $\mathbf{x}_\mathrm{A} \gets \mathrm{closestPointOnFace}(f_p, q)$\;
        $\mathbf{n} \gets \mathbf{x}_q - \mathbf{x}_\mathrm{A}$\;
        $\phi(f_p, q) \gets \|\mathbf{n}\| - \xi$\;
        \If{$\phi(f_p, q) < 2r$}{
            $w_f \gets \sum_{v \in f}{(\mathrm{barycentric}_v w_v)}$\;
            $\mathbf{n}_k \gets \mathbf{n} / \|\mathbf{n}\|$\;
            $d_k \gets \frac{w_f}{w_p + w_f}\,\phi(f_p, q)$\; 
            $\phi_k \gets \phi(f_p, q)$\;
            $\mathcal{C} \gets \mathcal{C} \cup \{(d_k, \mathbf{n}_k, \phi_k)\}$\;
        }
    }
}

\Return{$\mathcal{C}$}\;
\end{algorithm}

\section{Integration into XPBD}

This section describes how the proposed CCFR formulation is integrated into the PBD/XPBD frameworks.
Non-penetration and contact responses are expressed as local equality
constraints, allowing the method to be seamlessly incorporated into
existing XPBD solvers.

\paragraph{Penetration-Free Constraints}
To ensure penetration-free motion within XPBD, non-penetration is formulated as a position-level equality constraint via projection onto the convex feasible region $\mathcal{K}$ as defined in Equation 3 in the paper. 
The constraint function is defined as the Euclidean distance from the updated position $\mathbf{x}'$ to its closest projection $\mathbf{x}^*$ on the boundary of $\mathcal{K}$:
\[
    C_c^{\text{col}}(\mathbf{x}') = \|\mathbf{x}^* - \mathbf{x}'\|.
\]
The corresponding gradient $\nabla C(\mathbf{x}')$ is given by:
\[
    \nabla C_c^{\text{col}}(\mathbf{x}') = \frac{\mathbf{x}^* - \mathbf{x}'}{\|\mathbf{x}^* - \mathbf{x}'\|}.
\]
Note that this gradient is well-defined whenever $\mathbf{x}' \notin \mathcal{K}$, corresponding to a state where the non-penetration constraint is violated.

\paragraph{Algorithm of XPBD}
The overall algorithm is outlined in Algorithm~\ref{algorithm:xpbd}.
First, the predicted positions $\tilde{x}$ are computed (line~\ref{line:xpbd-prediction}).
Subsequently, collision detection is executed to determine the convex constraints for each vertex (lines \ref{line:xpbd-broad-phase}--\ref{line:xpbd-narrow-phase}).
Following initialization (line~\ref{line:xpbd-init-x}), the positions $x$ are iteratively updated by projecting each vertex to satisfy the corresponding constraints via a linear solver (lines~\ref{line:xpbd-iteration-start}--\ref{line:xpbd-iteration-end}).
Unlike the standard XPBD formulation, the solver loop of this algorithm encapsulates the entire substep process, including broad and narrow phase collision detection as well as convex region projections. 
This design effectively handles secondary collisions while mitigating excessive restrictions imposed by the maximum allowable step $r$, enabling the intended advancement of the current substep.
After applying the standard XPBD constraints (lines~\ref{line:xpbd-loop-start}--\ref{line:xpbd-loop-end}), both the physically active response and collision resolution are treated as equality constraints within XPBD, using $C_c^{\text{resp}}$ for enforcing active contacts
(line~\ref{line:xpbd-contact}) and $C_c^{\text{col}}$ for collision resolution via convex projection (line~\ref{line:xpbd-project-convex}).
%
%
To prevent collisions with primitives not detected during the initial broad and narrow phases (e.g., secondary collisions), any vertex that exceeds the maximum allowable step $r$ from its initial position is projected back to satisfy this limit (line~\ref{line:xpbd-keep-circle}).
For the convergence check (line \ref{line:xpbd-evaluate-convergence}), if the current step size falls short of the expected advancement, further iterations are performed; otherwise, the loop terminates.
Finally, the positions and velocities for the next time step are updated using the Implicit Euler method (lines \ref{line:xpbd-update-pos}--\ref{line:xpbd-update-vel}).

\begin{algorithm}[!t]
\SetAlgoNoLine
\LinesNumbered
\caption{Our simulation loop at substep $n$}
\label{algorithm:xpbd}
\KwIn{$x \coloneqq (\mathbf{x}_0, \mathbf{x}_1, \ldots)$: current positions, 
    $v \coloneqq (\dot{\mathbf{x}}_0, \dot{\mathbf{x}}_1, \ldots)$: current velocities, 
    $\mathcal{C}_\mathrm{XPBD}$: a set of XPBD constraints,
    time step size $\Delta t = \Delta t_\mathrm{frame} / N_\mathrm{substeps}$.}
\KwOut{$x^{n+1}$: next positions, $v^{n+1}$: next velocities.}
\vspace{0.5em}
\hrule
\vspace{0.5em}

compute predictions $\tilde{x} \gets x + \Delta t v + \Delta t^2 g$\;\label{line:xpbd-prediction}
initialize multipliers $\lambda \gets \mathbf{0}$\;

\For{$i = 1$ \KwTo $i_\mathrm{solverIterations}$}{\label{line:xpbd-iteration-start}
    broad phase at $x$\;\label{line:xpbd-broad-phase}
    narrow phase for each vertex $x$\tcp*{Alg. \ref{alg:NarrowPhase_detail}.}\label{line:xpbd-narrow-phase}
    initialize solve $x \gets \tilde{x}$\;\label{line:xpbd-init-x}
    \ForEach{constraint $c \in \label{line:xpbd-loop-start}\mathcal{C}_\mathrm{XPBD}$}{
        compute $\Delta \mathbf{\lambda}_c, \Delta x_c$\;
        update $x \gets x + \Delta x_c$\;
    }\label{line:xpbd-loop-end}
    \ForEach{$\mathbf{x}_k \in x$}{
    applyResponse$(\mathbf{x}_k)$\;\label{line:xpbd-contact}
    }
    \ForEach{$\mathbf{x}_k \in x$}{
    projectOntoConvexRegion$(\mathbf{x}_k)$\tcp*{Alg. \ref{alg:projectOntoConvexRegion}.}\label{line:xpbd-project-convex}
    }
    
    Project the vertex to enforce the maximum allowable step size $r$\;\label{line:xpbd-keep-circle}
    \If{evaluateConvergence}{\label{line:xpbd-evaluate-convergence}
        break\;
    }
}\label{line:xpbd-iteration-end}

update positions $x^{n+1} \gets x$\;\label{line:xpbd-update-pos}
update velocities $v^{n+1} \gets (x - x^n)/\Delta t$\;\label{line:xpbd-update-vel}
\end{algorithm}

\section{Details of Simulation Parameters}

Table~\ref{tab:parameters} summarizes the main parameters used throughout the simulations.
The search radius $r$ defines the maximum displacement range considered when constructing local feasible regions.
The offset parameter $\xi$ controls the conservative separation margin used for non-penetration handling, while the repulsion range $\delta$ determines the activation distance of the soft response model.

Unless otherwise stated, the same parameter settings were used consistently across each simulation category without extensive per-scene tuning.

\begin{table}[H]
    \centering
    \caption{Simulation parameters used in the experiments.} 
    \label{tab:parameters}
    \resizebox{\columnwidth}{!}{%
    \begin{tabular}{lrrrr}
        \toprule
        & search radius $r$  & offset $\xi$ & repulsion range $\delta$ \\
        \midrule      


        
        Twisting cloth & 0.005 & 0.01 & 0.0005  \\




        Soft Wires & 0.005 & 0.004 & 0.0008 \\

        Hair & 0.1 & 0.04 & 0.02 \\ 

        Particles & 0.005 & 0.005 & 0.004\\ 

        Codimensional & 0.4 & 0.06 & 0.01 \\ 
        \bottomrule
    \end{tabular}%
    }
\end{table}

\end{document}